# Wafer-scale hybrid molecular beam epitaxy of BaTiO$_3$ and SrTiO$_3$ on silicon

Xiaodong Tian[1,#], Yan Lin[2,#], Hanbin Gao[3,#], Han Yu[1], Yunpeng Ma[1], Ruiqi Liang[1], Changfu Chen[1], Wei Li[1], Chenguang Deng[1,*], Qiang Zheng[3], Qian Li[1,*]

[1]State Key Laboratory of New Ceramic Materials, School of Materials Science and Engineering, Tsinghua University, Beijing 100084, P. R. China

[2]School of Materials Science and Engineering, Peking University, Beijing, P. R. China

[3]CAS Key Laboratory of Standardization and Measurement for Nanotechnology, National Center for Nanoscience and Technology, Beijing 100190, China

[#]These authors contributed equally.

*Emails: dengchenguang@btifii.org.cn (C.D.); qianli_mse@tsinghua.edu.cn (Q.L.)

## ABSTRACT

The integration of epitaxial barium titanate (BTO) on silicon represents a highly promising pathway for next-generation, energy-efficient photonic integrated circuits due to BTO's exceptionally high Pockels coefficients. However, the scalable epitaxy of BTO on Si remains challenged by complicated stoichiometry control and slow growth rates. In this work, we demonstrate the continuous, wafer-scale growth of high-quality BTO films on SrTiO$_3$ (STO)-buffered 4-inch Si(001) wafers using a fully hybrid molecular beam epitaxy (hMBE) approach. By utilizing titanium tetraisopropoxide as a titanium precursor, we achieve a self-regulating, adsorption-controlled layer-by-layer growth at rates exceeding 75 nm h$^{-1}$, while maintaining an atomically sharp and structurally coherent BTO/STO interface. We systematically compare the structural, ferroelectric, and electro-optic (EO) properties of fully hMBE-grown BTO with those deposited via pulsed laser deposition (PLD) on identical STO/Si templates. While both techniques yield high-quality c-domain dominated films, the optimized hMBE-grown BTO exhibits superior crystallinity and a larger effective EO coefficient of 248 pm V$^{-1}$, surpassing that of the PLD-grown films (220 pm V$^{-1}$). These results highlight the advantages of the fully hMBE approach as a scalable, deterministic, and high-performance materials platform for wafer-scale integrated ferroelectric photonics.

# 1. INTRODUCTION

The relentless push for higher bandwidth and minimal energy dissipation in next-generation optical interconnects has exposed the intrinsic limitations of current photonic platforms [1–3]. Despite its dominance of large-scale integration, silicon is a centrosymmetric crystal structurally devoid of a linear electro-optic (Pockels) response. Consequently, high-speed light modulation in silicon relies primarily on carrier-dispersion effects, which inevitably suffer from issues such as free-carrier absorption and signal distortion [4–6]. While thin-film lithium niobate (LN) offers excellent linear modulation, its Pockels coefficient of approximately 30 pm $V^{-1}$ imposes a lower bound on device miniaturization and driving voltage reduction [7–9]. To break through these bottlenecks, epitaxial barium titanate (BTO) has emerged as a compelling material to functionally complement silicon photonics [10–12]. By combining a Pockels coefficient over an order of magnitude higher than that of LN with direct epitaxial integration on Si substrates, BTO provides a promising route toward compact, energy-efficient photonic integrated circuits (PICs) [13].

However, the epitaxial growth of high-quality BTO on Si presents specific technical challenges. Even when exploiting a 45° in-plane lattice rotation, the oxidizing environment required for perovskite synthesis typically forms an amorphous $SiO_2$ layer, disrupting the essential crystalline registry [14–16]. To address this, oxide molecular beam epitaxy (MBE) utilizes a sub-monolayer Sr reconstruction to passivate the Si surface, subsequently growing a $SrTiO_3$ (STO) buffer layer to serve as a compatible epitaxial template [17–19]. Nevertheless, continuing the BTO growth via conventional oxide MBE introduces practical limitations for scalable manufacturing. The low oxygen pressure characteristic of standard MBE processes generally leads to elevated oxygen vacancy concentrations. Furthermore, Ti metal sources exhibit low and unstable evaporation fluxes from a solid effusion cell, which restricts the standard MBE process to slow growth rates [20].

To address these scalability limitations, hybrid MBE (hMBE) using titanium tetraisopropoxide (TTIP) as the Ti source offers a distinct mechanistic pathway [21,22]. TTIP's specific decomposition kinetics create a self-regulated, adsorption-controlled growth window that eliminates the strict flux-matching constraints of conventional MBE, thereby enabling higher growth rates [23,24]. Furthermore, upon thermal decomposition at the heated substrate, TTIP acts as both the titanium source and a local oxygen supplier [25]. This intrinsic oxygen supply fulfills the lattice requirement while maintaining a sufficiently mild oxidizing environment to preserve the STO/Si interface [26]. By reducing the reliance on an external oxygen flux, the hMBE approach affords substantial process robustness for high-quality oxide epitaxy on silicon.

While this TTIP-driven approach provides an flexible framework for oxide epitaxy, the continuous synthesis of both STO buffer layers and BTO films entirely via hMBE on large-scale silicon wafers has yet to be demonstrated [27–31]. In this work, we realize this fully hMBE-based integration of BTO films on STO-buffered 4-inch Si(001) wafers. By utilizing TTIP as the Ti precursor, we achieved a BTO growth rate of ~75 nm h$^{-1}$ while strictly maintaining an adsorption-controlled, layer-by-layer growth mode. Structural characterizations confirm the pristine monocrystalline nature, atomically sharp interfaces, and well-defined atomic terraces of these hMBE-grown heterostructures. We additionally deposited BTO films on these virtual substrates using pulsed laser deposition (PLD) and radio-frequency (RF) magnetron sputtering. This comparative approach validates the structural robustness of the underlying hMBE-grown STO templates while simultaneously benchmarking the domain states and electro-optic (EO) properties of the BTO films grown by these different techniques.

Comprehensive EO characterizations illuminate the distinct physical mechanisms underlying these growth techniques. The PLD-grown BTO films on the hMBE-grown STO templates exhibit a high effective EO coefficient of 220 pm V$^{-1}$, confirming the excellent crystalline quality of the underlying STO buffer. However, this substantial response is primarily driven by uncontrolled, defect-induced strain relaxation, a stochastic process that inherently limits device-to-device reproducibility [32,33]. The fully hMBE-grown BTO films not only exhibit superior crystallinity and low leakage (with a high resistivity of $2.2\times10^4$ Ω·cm), but also a larger effective EO coefficient of 248 pm V$^{-1}$. Although both the PLD and hMBE films are predominantly c-domain oriented, this result indicates that the highly controllable, layer-by-layer growth mode of hMBE can achieve EO performance exceeding that of defect-mediated PLD films. Furthermore, while the TTIP-driven process provides a broad self-regulated growth window, we observe that the macroscopic EO properties remain sensitive to the specific TTIP flux ratio, likely due to subtle variations in defect concentrations as previously reported [31]. Crucially, the deterministic nature of the hMBE growth mode suggests substantial room for future property enhancement and optimization. For example, by tuning STO buffer thickness that affects strain transfer, successive BTO films can be engineered to favor an in-plane a-domain orientation, potentially unlocking even higher EO responses. Therefore, this work demonstrates the successful continuous wafer-scale epitaxy of high-quality BTO and STO films on silicon, enabling next-generation, high-performance EO modulators for optical interconnect applications.

## 2. EXPERIMENT

### 2.1 Thin-film epitaxy

In this work, hybrid MBE was implemented by modifying a commercial 4-inch MBE system (SVT Associates). The growth chamber was evacuated by a turbomolecular pump and a cryopump, together with a liquid-nitrogen-cooled cryopanel, resulting in a base pressure of $\sim 2\times 10^{-10}$ torr when the cryopanel was filled. Elemental Sr and Ba beams were supplied from standard effusion cells charged with high-purity source metals. Their beam-equivalent pressures (BEPs), measured using an ion gauge at the growth position, were varied from $1\times 10^{-8}$ torr to $1\times 10^{-7}$ torr by adjusting the cell temperatures. The Ti-containing metalorganic precursor was introduced by evaporating TTIP through a customized bubbler and injector source. The TTIP flux was controlled using a variable leak valve in combination with a temperature-controlled heating bath for the bubbler, and was typically maintained at 10–50 times the BEPs of Sr and Ba. Atomic oxygen was supplied using a RF plasma source operated at 300 W with an oxygen flow rate of up to 2 sccm.

High-resistivity Si(001) substrates, either as $10\times 10$ mm$^2$ chips or 4-inch wafers, were first rinsed in buffered oxide etch (BOE) solution to remove native SiO$_2$ layers and then quickly loaded into the MBE chamber. The substrates were radiatively heated up to 900 °C using a pyrolytic boron nitride (PBN) heater, with the temperature calibrated by an infrared pyrometer. Film growth was monitored in situ by reflection high-energy electron diffraction (RHEED, Staib). For wafer-scale growth, the 4-inch wafers were continuously rotated at 30 rpm to improve thickness and composition uniformity. Epitaxial STO thin films were first deposited to form STO/Si virtual substrates, followed by BTO film epitaxy either in-line or after air exposure/reloading. The detailed epitaxy growth route on Si is discussed in the Results section.

For comparison, BTO films were also deposited on STO/Si virtual substrates by pulsed laser deposition (PLD) and RF sputtering. For the PLD growth, a BTO ceramic target was ablated using a 248 nm KrF excimer laser with a fluence of 1 J cm$^{-2}$ and a repetition rate of 2 Hz. The substrates were maintained at 680 °C in an oxygen pressure of 5 Pa during deposition. For RF sputtering, a BTO ceramic target was sputtered by Ar plasma at a RF power of 75 W, while the substrates were heated to 700 °C in an Ar/$O_2$ mixed atmosphere with a ratio of 4:1 and a total pressure of 1.5 Pa. In both cases, the STO/Si virtual substrates used for BTO deposition were cut from the same hybrid MBE-grown wafer to ensure a meaningful comparison among the different growth techniques. Further details of the deposition conditions are provided in our previous reports [32,33].

**2.2 Structural characterizations**

The crystal structures of the films were analyzed by X-ray diffraction (XRD) using a diffractometer (Rigaku Smartlab) equipped with a 9 kW rotating Cu anode X-ray source and a HyPix-3000 high-resolution detector. For wafer-scale mapping, the 4-inch wafer was measured on a grid of 50 points using a motorized $xy$ stage while the diffraction optics remained fixed. The obtained XRD patterns were subsequently fitted to extract the diffraction peak positions and related structural information. Surface morphology and ferroelectric domain structures were examined using an atomic force microscope (AFM, Asylum Research MFP-3D). Piezoresponse force microscopy (PFM) images were collected in both vertical and lateral response channels.

For scanning transmission electron microscopy (STEM) characterization, cross-sectional specimens of the STO/Si and BTO/STO/Si films were prepared along the Si [110] zone axis using focused ion beam (FIB) in a dual-beam microscope (Zeiss Crossbeam). STEM was performed on a double-aberration-corrected transmission electron microscope (Spectra 300, Thermo Fisher Scientific) at an accelerating voltage of 300 kV. High-angle annular dark-field (HAADF) images were collected under a probe convergence angle of 25 mrad and with a detector collection angle of 49–200 mrad. Elemental distributions were analyzed by energy-dispersive X-ray spectroscopy (EDS), with maps collected simultaneously using a Super-X detector.

Time-of-flight secondary ion mass spectrometry (ToF-SIMS) measurements were performed on a reflection-type ToF-SIMS system (M6, IONTOF GmbH) equipped with a 30 keV $Bi^+$ liquid metal ion gun and a Cs ion beam sputter gun.

**2.3 Optical property measurements**

Second harmonic generation (SHG) measurements were conducted using a custom-built laser-scanning microscopy system[34–37]. The fundamental excitation ($1\omega$) was provided by a Ti:sapphire femtosecond laser (MaiTai, Spectra-Physics) operating at a central wavelength of 800 nm, with a pulse duration of 35 fs and a typical power of 50 mW. The $1\omega$ beam was directed and focused onto the sample surface using a convex lens (numerical aperture ~0.5). The sample stage was fixed at a 45° tilt angle relative to the incident beam, enabling the specularly reflected SHG ($2\omega$) signals to be efficiently collected along the detection optical axis. The polarization state of the incident beam was continuously tuned using a zero-order half-wave plate, while the generated $2\omega$ emission was analyzed by a Glan-Taylor prism polarizer. The resultant SHG signals were subsequently spectrally isolated using proper band-pass filters and recorded by a photomultiplier tube.

Spectroscopic ellipsometry measurements were performed on a phase-modulated ellipsometer (Uvisel Plus, HORIBA) over a spectral range of 350–2100 nm with a spot size of 100 μm. The fitting model consisted of a silicon substrate, a 23nm-thick STO layer (determined by HADDF measurement), an approximately 200–300-nm-thick BTO layer (Drude model), and the fitting spectral range was 400–2000 nm.

For electro-optic measurements, Pt coplanar electrodes were fabricated on the BTO films by standard photolithography, metal deposition, and lift-off, with an electrode gap of ~10 μm. The EO measurements were carried out using a home-designed apparatus employing a single-mode 1550 nm diode laser (Cnilaser TEM-F1550). The polarization state of the incident beam was adjusted using a Glan-Taylor prism together with a motorized half-wave plate, and the beam was then focused by an objective lens into the region between the electrodes. After transmission through the sample, the optical signal was analyzed by a motorized quarter-wave plate and a thin-film polarizer and recorded by an InGaAs detector. Electrical bias was applied through the coplanar electrodes in the form of a combined AC and DC field, and the corresponding phase modulation signal was extracted with a lock-in amplifier. For electro-optic loop measurements, a 5 V peak-to-peak AC modulation at 10 kHz was superimposed on a swept DC offset, and the field-dependent refractive-index change ($\Delta n$) was recorded accordingly. To determine the linear electro-optic response, the effective EO coefficient $r_{eff}$ was obtained from the slope of $\Delta n$ as a function of the applied AC electric field under a saturated DC bias. The film thicknesses obtained by spectroscopic ellipsometry were used for calculating the $r_{eff}$.

## 3. RESULTS AND DISCUSSION

### 3.1 Epitaxy of SrTiO$_3$ on Si

Figure 1a presents the overall hMBE growth procedure. To prevent oxidation of the Si interface during subsequent growth, the classic Sr passivation strategy was employed. Specifically, 1/2 monolayer (ML) of Sr was deposited and subsequently annealed at 720 °C to enhance Sr surface diffusion. As shown in Figure 1b, compared to the RHEED pattern of the pristine Si substrate, the annealed surface exhibited a characteristic 2×1 reconstruction, indicating the successful formation of a Zintl phase template. Following surface passivation, 12 ML of STO was co-deposited with the Sr and TTIP sources at a low temperature of 400 °C. This step resulted in the formation of an amorphous STO layer. The sample was then annealed at 600 °C to drive solid-phase epitaxy. After annealing, the sample exhibited a streaky RHEED pattern, confirming an ideal 2D surface morphology. Furthermore, distinct half-order reflections were observed between the fundamental streaks along the <100> azimuth directions. This feature not only confirms the formation of a highly crystalline, stoichiometric STO film but also indicates the presence of a well-ordered $TiO_2$-terminated surface. Notably, no background oxygen was introduced during this stage, demonstrating that the TTIP precursor alone supplies sufficient lattice oxygen for the epitaxial growth of STO. Following the crystallization of the buffer layer, the substrate temperature was further elevated to 800 °C. Atomic oxygen was then introduced to facilitate the high-temperature epitaxial growth of STO (or BTO) to the targeted thickness. The in-situ RHEED pattern of the STO film after high-temperature growth was exceptionally sharp with a clean background, exhibiting no amorphous scattering. Importantly, it preserved the characteristic $TiO_2$-termination reconstruction streaks, indicating sustained high crystallinity and stoichiometric control throughout the growth process.

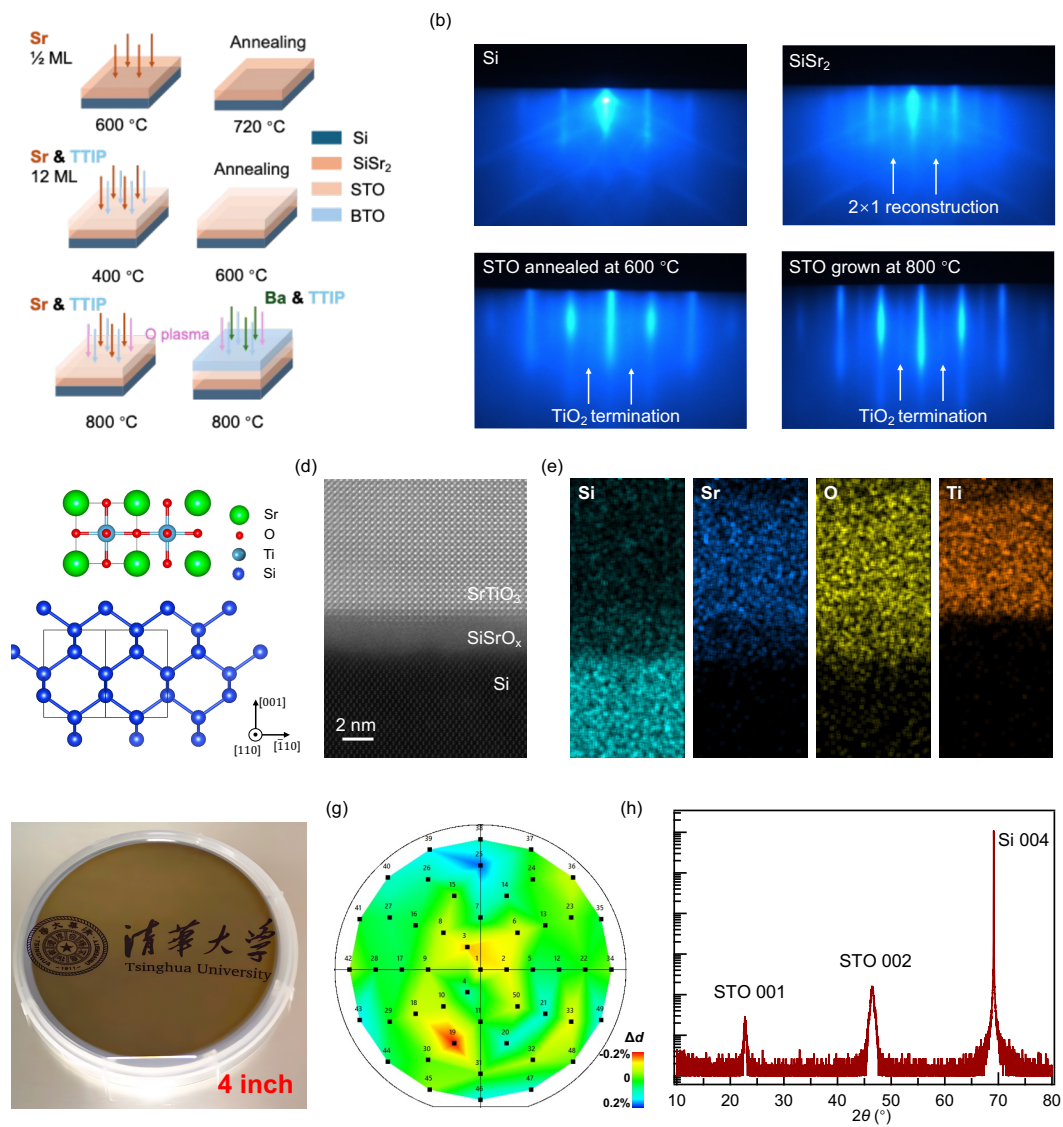

**Fig. 1** (a) Schematic of the developed growth route with typical process parameters. (b) Representative RHEED patterns recorded at various growth stages. (c) Illustration of the epitaxial relationship between STO and Si lattices. (d) Atomically resolved HAADF images for a STO/Si interface region and (e) corresponding EDS mapping results. (f) Photograph of a 4-inch 23 nm STO/Si wafer, with Tsinghua University Logo mirrored on the surface. (g) XRD mapping distribution of the (001) lattice spacing, relative to 3.911 Å, across the 4-inch wafer, and (h) $\theta-2\theta$ scan pattern for the center position, plotted in logarithmic intensity scale.

Figure 1c presents a cross-sectional schematic of the STO/Si heterostructure. Due to the lattice mismatch between STO ($a \approx 3.905$ Å) and Si ($a \approx 5.43$ Å), the STO lattice exhibits a 45° in-plane rotation relative to the Si substrate to minimize interfacial strain. Therefore, when viewed along the Si [110] zone axis, the STO lattice is projected along its [100] direction. This crystallographic model provides a reference for the cross-sectional HAADF-STEM image in Figure 1d. An amorphous interfacial layer is present between the Si substrate and the crystalline STO film. This layer forms during the 800 °C growth stage as oxygen diffuses through the STO lattice to the underlying Si surface. Despite the formation of this amorphous interface, the overlying STO film maintains its initial epitaxial orientation. This indicates that the epitaxial registry was established by the Zintl template and the solid-phase crystallized buffer layer prior to the high-temperature oxidation, enabling the continued epitaxial growth of the STO. As shown in Figure 1e, EDS mapping reveals that the amorphous interfacial layer consists primarily of Si, Sr, and O. The absence of Ti in this region indicates its confinement within the perovskite lattice during oxidation, whereas Sr forms a silicate-like layer. Furthermore, the weak Si signal detected throughout the STO film is an artifact arising from the EDS spectral overlap of Si and Sr, rather than actual atomic diffusion.

To evaluate the macroscopic uniformity of the STO film at the wafer scale, Figure 1f presents a photograph of a 4-inch STO/Si wafer with a 23 nm-thick STO film. The surface appears homogeneous, and the absence of interference color fringes indicates a uniform film thickness across the wafer. The crystalline quality and epitaxial orientation were characterized by XRD. A representative $\theta-2\theta$ scan (10°–80°) collected from the wafer center (Figure 1h) is plotted on a logarithmic intensity scale. The diffractogram exclusively displays the STO 00$l$ diffraction peaks, including the fundamental 001 and higher-order 002 peaks, alongside the Si 004 substrate peak. No secondary phases or misoriented grain peaks are detected, confirming phase-pure epitaxial growth. An XRD mapping of the STO (001) lattice spacing was performed (Figure 1g). Referenced to 3.911 Å, the lattice spacing is highly uniform across the entire wafer, with local deviations ($\Delta d/d$) strictly bounded within ±0.2%. Importantly, the XRD spectra from all mapping points consistently match the representative profile in Figure 1h, showing a complete absence of impurity phases or non-epitaxial peaks. This confirms the single-crystal epitaxial nature of the STO film across the entire 4-inch wafer.

### 3.2 Epitaxy of BaTiO$_3$ on SrTiO$_3$/Si

Building upon the structural stability of the STO/Si template, epitaxial BTO films were subsequently grown at 800 °C. In the hMBE process, the high volatility of the TTIP precursor provides an adsorption-controlled growth window. Within this thermodynamic regime, excess TTIP desorbs from the surface, rendering the growth rate governed by the Ba flux. This self-limiting mechanism allows for sustained epitaxial growth over a broad range of precursor flux ratios. For instance, two representative hMBE samples grown with TTIP/Ba flux ratios of 38:1 (hMBE_1) and 32:1 (hMBE_2) maintained sharp, streaky RHEED patterns throughout the deposition process. Notably, this hMBE process affords a growth rate of 70–100 nm h$^{-1}$ (e.g., ~75 nm h$^{-1}$ for the 220 nm-thick hMBE_2 film). This rate significantly outpaces conventional oxide MBE processes (typically < 20 nm h$^{-1}$).

To demonstrate the universal applicability of the STO/Si pseudo-substrate, BTO films were also deposited using PLD and RF sputtering. Figure 2b presents the RHEED patterns of the STO/Si template before and after the PLD growth of BTO. The preservation of sharp diffraction streaks confirms the layer growth mode of PLD. The crystalline quality and phase purity of the BTO films across all growth methods were evaluated using wide-range XRD $\theta-2\theta$ scans. As shown in Figure 2a, all four samples exclusively display the fundamental (001) diffraction peaks or their higher-order reflections for BTO, STO, and Si. The absence of impurity phases and non-epitaxial peaks indicates that the STO/Si template supports successful epitaxial integration across different deposition techniques. The inset of Figure 2a presents the BTO (002) peaks plotted on a linear scale. The hMBE_1 film exhibits an out-of-plane lattice parameter closely matching the bulk BTO a-axis. In contrast, the other samples display out-of-plane lattice parameters intermediate between the bulk a- and c-axes. Figure 2d further presents the reciprocal space mapping (RSM) of the hMBE_2 film. The extracted in-plane and out-of-plane lattice parameters are a = 3.993 Å and c = 4.017 Å, respectively, indicating a state closer to the bulk c-domain.

Based on the hMBE_2 process, a 4-inch BTO/STO/Si wafer was continuously grown in-line with a thinner STO buffer layer (~8 nm). Figure 2c presents the XRD mapping of the out-of-plane lattice spacing. Reference to a nominal lattice parameter of 3.996 Å, the map reveals a relatively uniform distribution, with $\Delta d/d$ deviation bounded within ±0.2%. In comparison with the STO/Si wafer (Figure 1g), the observed ring-shaped feature suggests a moderate non-smooth flux profile from the Ba effusion cell. Notably, despite using identical growth parameters, this 4-inch BTO film exhibits a reduced out-of-plane lattice parameter (3.996 Å) closer to the bulk a-axis. As reported in previous literature [38,39], this shift occurs because the thinner STO layer allows the biaxial tensile strain induced by the thermal expansion mismatch with the Si substrate to dominate.

To evaluate the depth-dependent chemical composition, ToF-SIMS was performed for the hMBE_2 film (Figure 2e). The carbon (C) signal exhibits a strong initial peak at the film surface, commonly caused by the physical adsorption of airborne $CO_2$ upon ambient air exposure. Within the BTO film, the C intensity drops drastically, stabilizing at a level only slightly above the baseline recorded in the underlying Si substrate. Since the pristine Si substrate is inherently carbon-free, this low baseline represents the instrumental background limit. This comparison confirms negligible carbon incorporation from the TTIP precursor. Furthermore, apparent intensity spikes are observed for multiple elements at the interfacial region. This is an artifact caused by the matrix effect in SIMS, where the abrupt change in the chemical environment at the oxide/semiconductor boundary alters the secondary ion yield. An anomalous Ba peak appears at the deep STO/Si interface. Since Ba was not introduced during the STO buffer layer growth, this signal is predominantly attributed to a knock-on artifact, wherein heavy Ba atoms from the top layer are physically driven into the substrate by the primary sputtering ion beam.

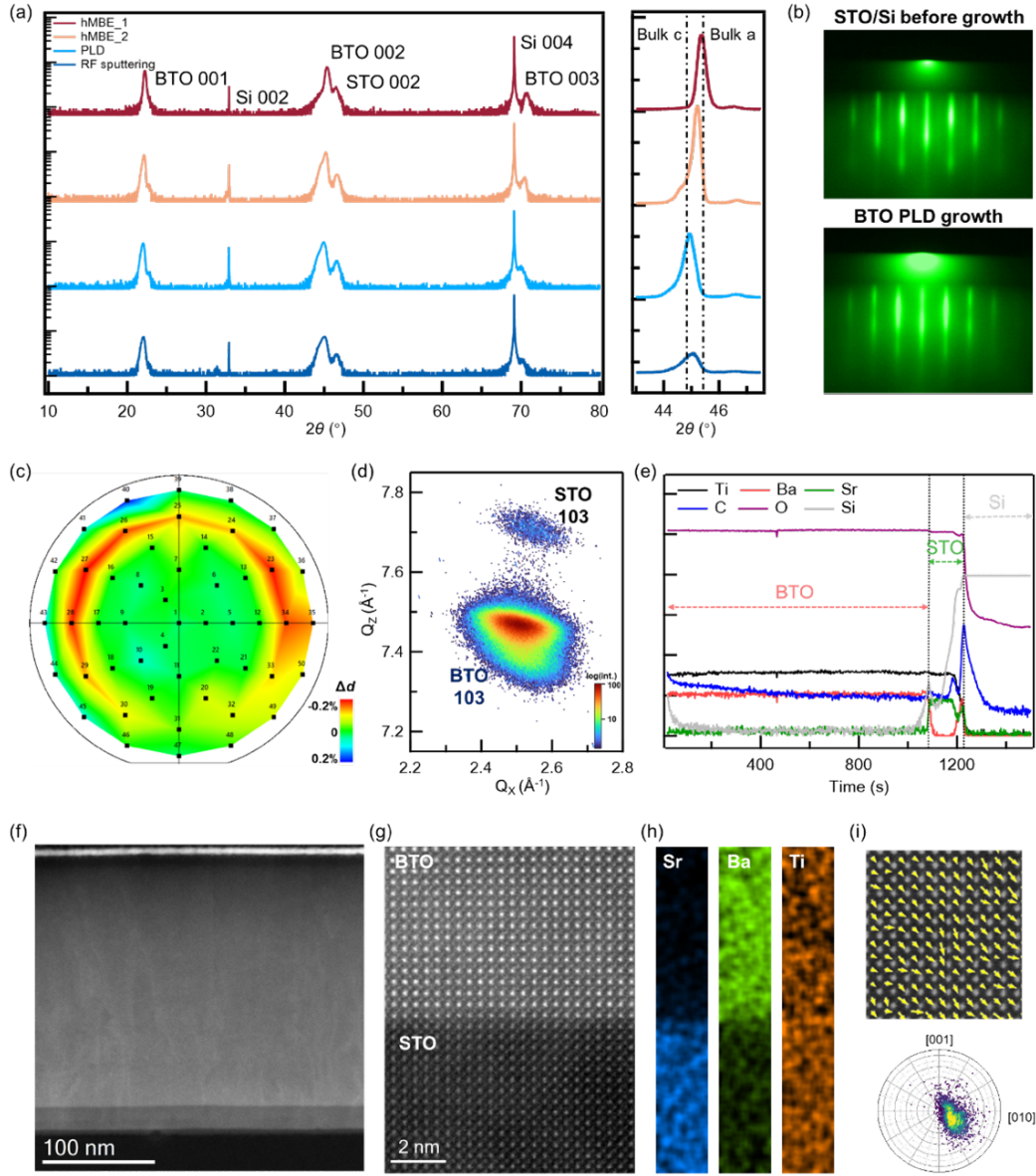

**Fig. 2** (a) Comparison of θ–2θ scan patterns for the BTO films grown via hybrid MBE, PLD and sputtering techniques on STO/Si substrates cut from the same original wafer. The *left* inset is a zoom-in view of BTO 002 peaks plotted in a linear intensity scale. (b) RHEED patterns recorded during the PLD growth. (c) XRD mapping distribution of the (001) lattice spacing, relative to 3.996 Å, across a 4-inch ~300 nm BTO/8 nm STO/Si wafer. (d) Reciprocal space mapping (RSM) around the asymmetric 103 reflections of BTO and STO. (e) Depth profiles of the chemical compositions obtained via Tof-SIMS for the hMBE_2 film. (f) Large field-of-view HAADF-STEM image of the BTO/STO/Si heterostructure. (g) High-magnification HAADF-STEM image of the BTO/STO interface. (h) Corresponding EDS elemental mappings for Sr, Ba, and Ti. (i) Selected-area polarization mapping and its corresponding statistical analysis.

To directly visualize the microstructural quality and epitaxial registry, HAADF-STEM was performed for the hMBE_2 film. As shown in the large field-of-view image (Figure 2f), the BTO film exhibits uniform contrast without any observable threading dislocations, planar defects, or secondary phase inclusions, corroborating the excellent macroscopic crystalline quality. The atomic structure of the BTO/STO interface is further resolved in the high-magnification HAADF-STEM image (Figure 2g). The interface is atomically sharp and structurally coherent. Notably, no significant interfacial misfit dislocations were observed during extensive cross-sectional examinations across multiple regions. This indicates that the epitaxial lattice registry is well-preserved without obvious plastic strain relaxation, maintaining a highly pristine atomic template. Figure 2h displays the corresponding EDS elemental mapping for Sr, Ba, and Ti. The mappings visually confirm the distinct chemical stratification of the BTO and STO layers. While a faint Ba signal appears to be distributed within the STO layer, this is a well-documented EDS deconvolution artifact caused by the severe spectral overlap between the Ba L-lines (~4.46 keV) and the Ti K-lines (~4.51 keV). As definitively established by the highly sensitive ToF-SIMS depth profiling, the STO buffer layer is essentially Ba-free. Thus, the complementary structural and chemical analyses verify the formation of a structurally coherent and chemically abrupt BTO/STO interface. The selected-area polarization mapping (Figure 2i) and its corresponding large-area statistical analysis reveal a mixed a/c-domain configuration dominated by out-of-plane c-domains.

**3.3 Ferroelectric domain structures**

To comprehensively evaluate the surface morphology and the local-to-macroscopic ferroelectric properties of the BTO films, we performed AFM/PFM and SHG measurements. Figure 3 presents a comparative analysis of the three samples: two fully hMBE-grown BTO films (hMBE_1 and hMBE_2) and the PLD-grown BTO film on an hMBE-STO template. As shown in the AFM images (Figure 3a–c), all three BTO films exhibit exceptionally smooth surfaces. The height variations across the measured areas are strictly confined within ±400 pm, corresponding to roughly a single unit-cell step height. Notably, clear atomic step-terrace structures are distinctly resolved on the surfaces of both hMBE_1 and hMBE_2. These well-defined atomic terraces unambiguously confirm that an ideal, adsorption-controlled layer-by-layer growth mode is maintained throughout the hMBE process. While the PLD-grown film also exhibits sub-nanometer roughness, it lacks these distinct terrace features, highlighting the mechanistic difference between the thermodynamically governed hMBE and the more energetic, nonequilibrium PLD process.

Figure 3d–f presents the PFM images in four channels: vertical amplitude, lateral amplitude, vertical phase, and lateral phase. The PFM images reveal a distinct evolution of the static domain states across the three samples. For the hMBE_1 film (Figure 3d), the PFM response is characterized by a horizontal boundary separating the region into two parts: the upper region exhibits prominent lateral amplitude (in-plane polarization), while the lower region is dominated by vertical amplitude (out-of-plane polarization). Accompanied by abrupt phase inversions across this boundary, this configuration represents classic 90° domain walls separating *a*- and *c*-domains. In contrast, the hMBE_2 film (Fig. 3e), grown with a reduced TTIP flux ratio, exhibits a relatively uniform vertical amplitude with slight phase fluctuations, while the lateral amplitude displays finer, more localized intensity variations accompanied by 90° phase flips. This fragmentation trend is further amplified in the PLD-grown film (Fig. 3f), where the vertical phase is highly uniform, and the lateral signal clearly resolves fine-grained domain structures with frequent phase flips. Overall, the progression demonstrates a morphological evolution where the ferroelectric domains become increasingly

fragmented, and the out-of-plane *c*-domains gradually establish dominance, in line with the XRD measurement results.

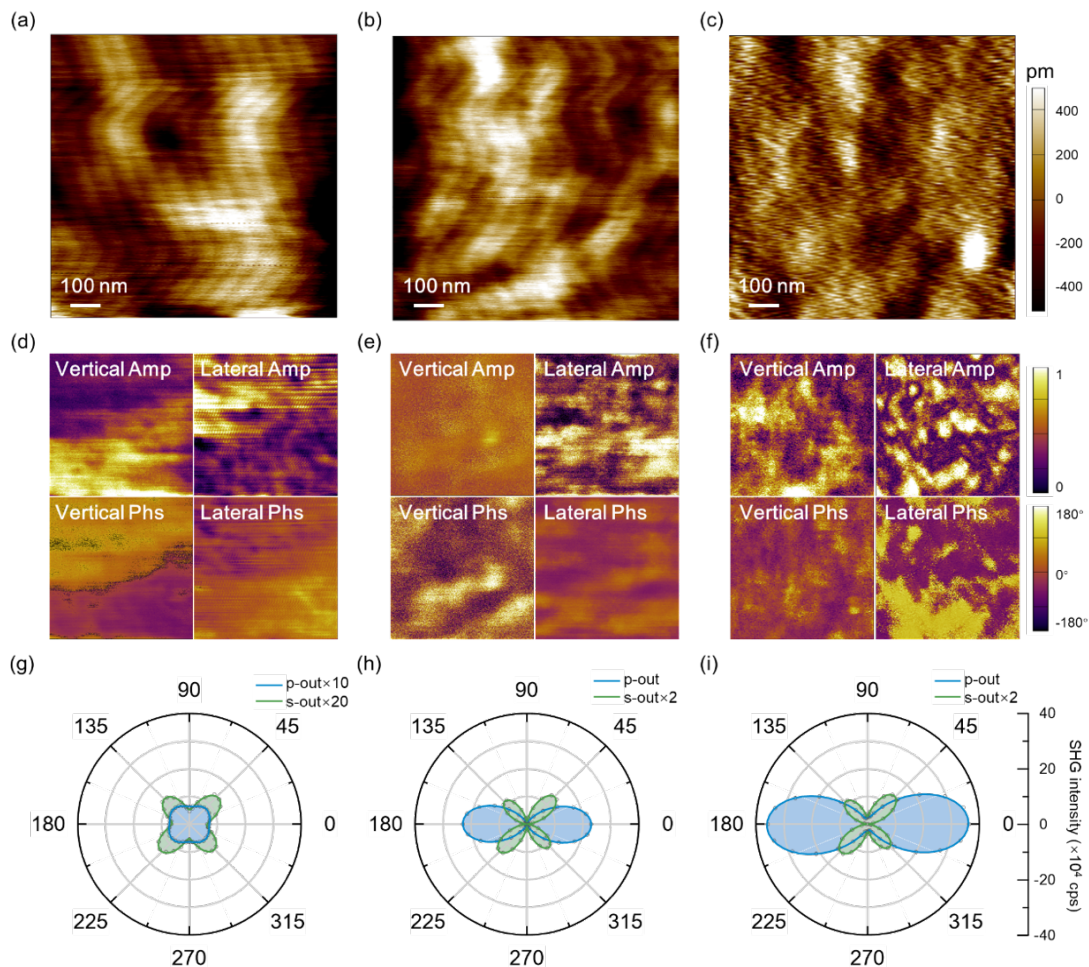

**Fig. 3** (a–c) AFM topography images of the fully hMBE-grown BTO films (hMBE_1 and hMBE_2) and the PLD-grown BTO film on an hMBE-STO template. (d–f) PFM measurements of the corresponding samples. Each panel is divided into four quadrants: vertical amplitude (*top-left*), lateral amplitude (*top-right*), vertical phase (*bottom-left*), and lateral phase (*bottom-right*). (g–i) SHG polarimetry plots measured as a function of the incident fundamental beam polarization angle under the s-out and p-out detection configurations.

To bridge the local domain configurations with macroscopic optical characteristics, SHG polarimetry was performed by rotating the incident fundamental beam polarization while keeping the output polarization fixed in either the *s*-out or *p*-out configuration (Fig. 3g–i). In the *s*-out configuration, all three samples commonly exhibit a clear four-fold symmetry with maximum intensities at 45°, consistent with the fundamental tetragonal *4mm* symmetry of epitaxial BTO. However, the *p*-out configurations and overall SHG intensities reveal stark differences that directly map to their specific domain evolutions. The hMBE_1 film exhibits a nearly isotropic *p*-out response coupled with a very low overall SHG intensity, suggesting a weakly developed polar state with a pseudo-cubic structure. In contrast, both the hMBE_2 and PLD-grown films display a clear two-fold symmetry along 0° in the *p*-out configuration, confirming their transition into a *c*-domain dominated tetragonal state. Although both hMBE_2 and the PLD films share a similar *c*-domain dominant architecture, their SHG responses exhibit a clear progressive enhancement. Driven by the increasing tetragonality and *c*-domain dominance, the effective nonlinear optical coefficient $d_{33}$ is observed to gradually increase from hMBE_2 to the PLD-grown film. This systematic evolution perfectly traces the structural transition from an *a*-domain dominated pseudo-cubic state toward a highly robust, *c*-domain dominated tetragonal state, in line with the PFM and XRD measurement results.

### 3.4 Electro-optical properties

Building on the structural and polarization-sensitive characterizations above, we next examine the optical responses of the three BTO films. Figure 4a shows the optical constants extracted from spectroscopic ellipsometry. All three films exhibit similar refractive indices in the C-band telecommunication wavelength range, and these values are consistent with those typically reported for bulk $BaTiO_3$. Meanwhile, the extinction coefficient remains close to zero, indicating low optical absorption loss in the operating wavelength range.

Prior to the EO characterizations, the electrical resistivities of the films were determined to be in the range of 2.0–2.2 × 10⁴ Ω·cm, ensuring low leakage currents under for high lateral electric fields. Despite their similarly high quality, the three films exhibit markedly different electro-optic responses. As shown in Figure 4b, the PLD-grown film displays a pronounced bipolar hysteresis loop with a clear constriction and step-like feature near zero field. Such a loop shape can be understood in terms of the competition between electric-field-driven polarization reorientation and elastic restoring forces within the film. The dashed gray envelope in Figure 4b is a schematic reference for an ideal switching behavior in the absence of elastic clamping. In the actual films, part of the field-switched regions remains elastically constrained by the compressive strain associated with the STO buffer, forming a partially stable switched state. Under a strong in-plane electric field, these regions can be driven toward an in-plane configuration, accompanied by the buildup of elastic strain energy. Upon reducing the electric field toward zero, the electrostatic energy provided by the external field becomes insufficient to counterbalance the elastic restoring force, causing part of the switched domains to relax prematurely back toward the original c-oriented state. As a result, part of the ideal switching contribution is lost along the descending branch, leading to the constricted and step-like loop shape near zero field.

The behavior of the two hMBE-grown films further highlights the sensitivity of the electro-optic response to specific domain configurations. Although both films remain within the high-quality growth window and show nearly indistinguishable average structural characteristics (Figure 2), the effect of flux-ratio variation becomes evident in the polar-state-sensitive SHG response (Figure 3). This difference is further reflected in their distinct EO responses. Consistent with its cubic-like SHG response, the hMBE_1 film displays only a very slim EO loop with a small modulation amplitude and no obvious hysteresis, as shown in Fig. 4c. By contrast, the hMBE_2 film shows an SHG response much closer to that of the PLD-grown film, indicating the similar mixed a/c-domain configuration. Accordingly, hMBE_2 also exhibits a much larger bipolar electro-optic loop with a constriction near zero field, as shown in Figure 4d, closely resembling the behavior of the PLD-grown film. Such similarity suggests a common switching scenario involving partial field-induced c-to-a reorientation followed by elastic back-relaxation during field removal.

This difference is further reflected in the effective EO coefficient ($r_{eff}$) measurements, as shown in Fig. 4e. All films exhibit linear dependence on the AC electric field, attesting to Pockels effect responses. The extracted $r_{eff}$ is 32 pm V$^{-1}$ for hMBE1, much lower than the values of 248 pm V$^{-1}$ and 220 pm V$^{-1}$ obtained for hMBE_2 and the PLD-grown film, respectively. Despite their similar structural signatures, the pronounced difference in EO response between the two hMBE-grown films suggests that structural characterization alone is insufficient to fully capture the functional variation of these films. Taken together, these results indicate that the broad growth window of hMBE does not ensure equivalent device-relevant properties across different flux-ratio conditions, and that careful optimization of the Ba/Ti flux ratio remains essential for achieving strong electro-optic performance.

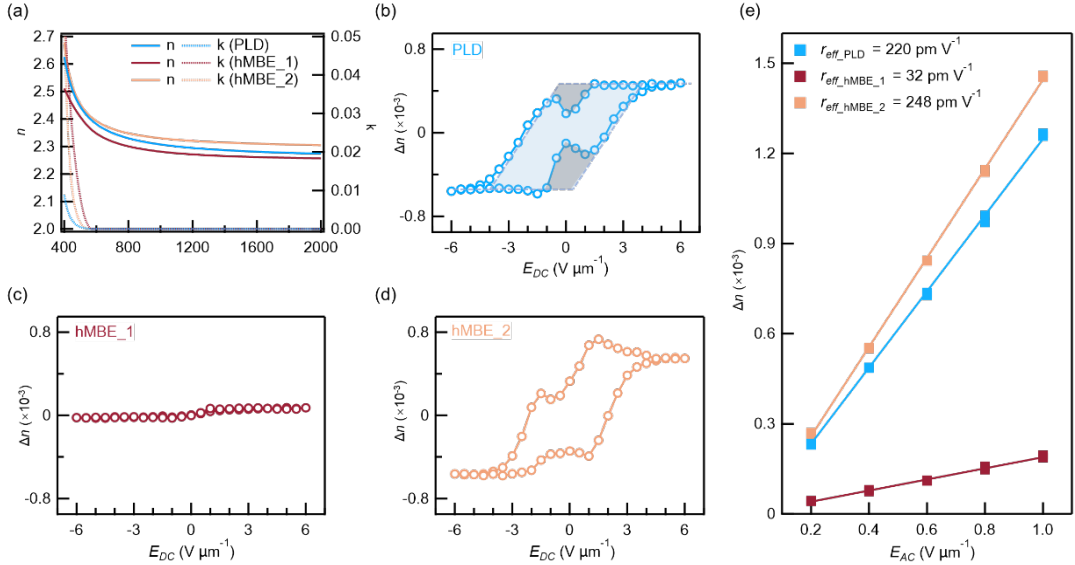

**Fig. 4** (a) Wavelength-dependent refractive index ($n$) and extinction coefficient ($k$) extracted from spectroscopic ellipsometry for the PLD- and hMBE-grown BTO films. (b–d) Refractive-index change ($\Delta n$) as a function of applied DC electric field for the (b) PLD-grown film, (c) hMBE_1 and (d) hMBE_2 thin films, showing their electro-optic hysteresis loops. The gray shaded envelope in (b) schematically represents the ideal EO switching response expected for a film dominated by out-of-plane ferroelectric domains in the absence of stress clamping. (e) Dependence of $\Delta n$ on the applied AC electric field under saturated DC bias, together with the extracted $r_{eff}$ for the three films.

## 4 CONCLUSIONS

In summary, we have realized the continuous, wafer-scale epitaxial growth of high-quality BTO/STO heterostructures on 4-inch Si(001) substrates via a fully hMBE process. The utilization of a TTIP precursor effectively circumvents the tight flux-matching and slow-growth limitations associated with conventional oxide MBE, achieving a fast growth rate in excess of ~75 nm h$^{-1}$ while strictly preventing amorphous disruption at the STO/Si interface. Structural, ferroelectric, and optical characterizations confirm that this thermodynamically governed, layer-by-layer process yields highly crystalline BTO films with distinct atomic terraces. Crucially, the fully hMBE-grown BTO films demonstrate an outstanding effective electro-optic coefficient of 248 pm V$^{-1}$. Our findings also highlight that achieving this strong performance requires careful optimization of the TTIP/Ba flux ratio, despite the broad self-regulated growth window afforded by hMBE. Ultimately, the continuous hMBE growth of silicon-epitaxial BTO/STO heterostructures establishes a scalable, deterministic, and versatile materials platform for integrated ferroelectric photonics, enabling compact and high-efficiency electro-optic devices while offering broad prospects for further performance enhancement through strain and interface engineering.

## ACKNOWLEDGEMENTS

This work was supported by Scientific Research Innovation Capability Support Project for Young Faculty under Grant No.ZYGXQNJSKYCXNLZCXM-M17, the Basic Science Center Project of National Natural Science Foundation of China (NSFC) under Grant No. 52388201, NSFC grants No. U24A2009, 12474087 and 52502142, Beijing Municipal Natural Science Foundation under Grant No. JQ24011 and Z240008, and the open research fund of Suzhou Laboratory (No.SZLAB-1508-2025-ZD016).

**Availability of data and materials**

The data that support the findings of this study are available from the corresponding author upon reasonable request.

**Competing interests**

The authors have no competing interests to declare that are relevant to the content of this article.